\documentclass[aps,twocolumn,superscriptaddress,prl,floatfix,british]{revtex4}
\usepackage{amssymb}
\usepackage{amsmath}
\usepackage{babel}
\usepackage[sort&compress]{natbib}
\usepackage{topcapt}

\usepackage{graphicx}
\graphicspath{{figs/}}
\usepackage[colorlinks,linkcolor=dred,urlcolor=dblue,citecolor=dgreen]{hyperref}
\usepackage{stmaryrd}
\usepackage{amsfonts}
\usepackage{mathrsfs}
\usepackage{txfonts}
\graphicspath{{figs/}}
\usepackage{diagbox}
\usepackage{color}
\usepackage{pdfpages}
\definecolor{dred}{rgb}{.8,0.2,.2}
\definecolor{ddred}{rgb}{.8,0.5,.5}
\definecolor{dblue}{rgb}{.2,0.2,.8}
\definecolor{dgreen}{rgb}{.2,0.5,.2}

\usepackage{multirow}

\newcommand{\Ref}[1]{(\ref{#1})}

\newcommand{\ket}[1]{| #1 \rangle}

\newcommand{\vecx}{\mathbf{x}}
\newcommand{\vecX}{\mathbf{X}}

\newcommand{\matH}{\mathbf{H}}
\newcommand{\matI}{\mathbf{I}}
\newcommand{\opH}{\mathbf{H}}

\def\be{\begin{eqnarray}}
\def\ee{\end{eqnarray}}

\usepackage{times}

\begin{document}
\title{A Full Quantum Eigensolver for Quantum Chemistry Simulations}

\author{Shijie Wei}
\affiliation{Beijing Academy of Quantum Information Sciences,  Beijing 100193, China}
\affiliation{State Key Laboratory of Low-Dimensional Quantum Physics and Department of Physics, Tsinghua University, Beijing 100084, China}

\author{Hang Li}
\affiliation{State Key Laboratory of Low-Dimensional Quantum Physics and Department of Physics, Tsinghua University, Beijing 100084, China}
\author{GuiLu Long}
\email{gllong@tsinghua.edu.cn}
\affiliation{State Key Laboratory of Low-Dimensional Quantum Physics and Department of Physics, Tsinghua University, Beijing 100084, China}
\affiliation{ Beijing National Research Center for Information Science and Technology and School of Information Tsinghua University, Beijing 100084, China}
\affiliation{Beijing Academy of Quantum Information Sciences,  Beijing 100193, China}
\affiliation{Frontier Science Center for Quantum Information, Beijing 100084, China}
\date{\today}
\maketitle
\textbf{ Quantum simulation of quantum chemistry is one of the most compelling applications of quantum computing. It is of particular importance in  areas ranging from materials science, biochemistry and condensed matter physics.  Here, we propose  a full quantum eigensolver (FQE) algorithm to calculate the molecular ground energies and electronic structures using  quantum gradient descent. Compared to existing classical-quantum hybrid methods such as variational quantum eigensolver (VQE), our method removes the classical optimizer and performs all the calculations on a quantum computer with faster convergence. The gradient descent iteration depth  has a favorable complexity that is logarithmically dependent on the system size and  inverse of the precision. Moreover, the FQE can be further simplified by exploiting perturbation theory for the calculations of intermediate matrix elements, and obtain results with a precision that satisfies the requirement of chemistry application. The full quantum eigensolver can be implemented on a near-term quantum computer. With the rapid development of quantum computing hardware, FQE provides an efficient and powerful  tool to solve  quantum chemistry problems.}
\section{introduction}
Quantum chemistry  studies chemical systems using quantum mechanics.  One primary focus of quantum chemistry is  the calculation of molecular energies and electronic structures of a chemical system which determine   its chemical properties. Molecular energies and electronic structures are calculated by solving the    Schr\"{o}dinger equation within chemical precision. However, the computational resources needed  scale exponentially with the system size on a classical computer, making the calculations in quantum chemistry intractable in high-dimension. 
     
Quantum computers, originally envisioned by Benioff, Manin and Feynman \cite{benioff1980computer,manin1980vychislimoe,feynman1982simulating}, have emerged as promising tools for tackling this challenge with polynomial overhead of computational resources. Efficient quantum simulations of chemistry systems promise breakthroughs in our knowledge for basic chemistry and  revolutionize research in new materials, pharmaceuticals, and industrial catalysts.

 The universal quantum simulation method \cite{lloyd1996universal} and the first quantum algorithm for simulating fermions \cite{abrams1997simulation}  have laid down the fundamental block of quantum chemistry simulation. Based on these techniques and quantum phase estimation  algorithm  \cite{kitaev1995quantum}, Aspuru-Guzik et al presented a quantum algorithm for preparing ground states undergoing an adiabatic evolution \cite{aspuru2005simulated}, and many theoretical and experimental works \cite{babbush2014adiabatic,feng2013experimental,lu2015experimental,babbush2015chemical,wei2016dualityopen,babbush2016exponentially,babbush2017exponentially,kassal2008polynomial,kivlichan2017bounding,toloui2013quantum,peruzzo2014variational,mcclean2016theory,mcclean2014exploiting,whitfield2011simulation,wecker2015progress,hastings2014improving, kyriienko2019quantum} have been developed since then.  In 2002, Somma et al. proposed a scalable quantum algorithm for the simulation of molecular electron dynamics via Jordan-Wigner transformation \cite{jordan1928pauli}. The Jordan-Wigner transformation directly maps the fermionic occupation state of a particular atomic orbital to a state of qubits, which enables the quantum simulation of chemical systems on a quantum computer. Then, the Bravyi-Kitaev  transformation \cite{bravyi2002fermionic,seeley2012bravyi,tranter2015b,
bravyi2017tapering, babbush2018low} encodes both locality of occupation and parity information onto the qubits, which is more efficient in operation complexity. In 2014, Peruzzo  et al developed the variational quantum eigensolver (VQE) \cite{peruzzo2014variational,yung2014transistor}, which finds a good variational approximation to the ground state of a given Hamiltonian for a particular choice of ansatz. Compared to quantum phase estimation and trotterization of the molecular Hamiltonian, VQE requires a lower number of controlled operations and shorter coherence time.  However, VQE is a classical and quantum hybrid algorithm, the optimizer is performed on a  classical machine.

Meanwhile, implementations of quantum chemistry simulation have been developing steadily. 
Studies in  present-day quantum computing hardware have been carried out,  such as  nuclear magnetic resonance system  \cite{du2010nmr, li2019quantum}, photonic system \cite{roushan2017chiral,lanyon2010towards,paesani2017experimental},   nitrogen-vacancy center system \cite{wang2015quantum} , trapped ion \cite{shen2017quantum,hempel2018quantum} and superconducting system \cite{o2016scalable,kandala2017hardware, ganzhorn2019gate}. Rapid  development in quantum computer hardware with even the claims of quantum supremacy,   greatly stimulates the expectation of its real applications. Quantum chemistry simualtion is considered as a real application in Noisy Intermediate-Scale Quantum (NISQ) computers \cite{mohseni2017commercialize,wecker2015progress,mueck2015quantum}. The FQE is an effort on this background. In FQE, not only calculation of Hamiltonian matrix part is done on quantum computer, but also the optimization by gradient descent is  performed on quantum computer.
FQE can be used in  near-term NISQ computers, and in future fault-tolerant large quantum computers.
\section{method}
\subsection{Preparing the Hamiltonian for Quantum Chemistry Simulation}
A  molecular  system, contains a collection of nuclear charges $Z_i$ and  electrons. The fundamental task of quantum  chemistry is to solve the eigenvalue problem of the molecular Hamiltonian. The eigenstates of the many-body  Hamiltonian determine the dynamics of the electrons as well as the properties of the molecule.  The corresponding  Hamiltonian of the system includes kinetic energies of nuclei and electrons, the Coulomb potentials of nuclei-electron, nuclei-nuclei,  electron-electron and it can be expressed in first quantization as
\begin{equation}
\label{eq:firstHam}
\begin{aligned}
H_o &= - \sum_i \frac{\nabla_{R_i}^2 }{2M_i} - \sum_i \frac{\nabla_{r_i}^2}{2} - \sum_{i,j} \frac{Z_i}{|R_i - r_j|} \\
&+ \sum_{i, j > i} \frac{Z_i Z_j}{|R_i - R_j|} + \sum_{i, j>i} \frac{1}{|r_i - r_j|},
\end{aligned} 
\end{equation}
in atomic units $(\hbar =1)$, where  $R_i, Z_i, M_i$ and $r_i$ are the positions, charges, masses of the nuclei and the positions of the electrons respectively.  Under the Born-Oppenheimer approximation which assumes the nuclei as a fixed classical point, this Hamiltonian is usually rewritten in the particle number representation in a chosen basis 
\begin{align}\label{eq:ferm_hamiltonian}
H= \sum_{ij} h_{ij} a^{\dagger}_i a_j+ \frac{1}{2} \sum_{ijkl} h_{ijkl} a^{\dagger}_i a^{\dagger}_j a_k a_l + \dots ,
\end{align}
where $\cdots$ denotes higher order interactions and  $a_i^\dagger$ and $a_j$ are the creation and annihilation operator of particle in orbital $i$ and $j$ respectively. The parameters $h_{ij}$ and $h_{ijkl}$ are the one-body and two-body  integrations in the chosen basis functions $\{\psi_i\}$. In Galerkin formulation, the scalar coefficients in Eq.~\eqref{eq:ferm_hamiltonian} can be calculated by

\begin{align}\label{eq:integrals}
h_{ij}&=\langle\psi_i|\left(-\frac{\nabla_i^2}{2}-\sum_{A}\frac{Z_A}{\left|r_i-R_A\right|}\right)|\psi_j \rangle\\\nonumber
h_{ijkl}&=\langle\psi_i\psi_j|\frac{1}{\left|r_i-r_j\right|}|\psi_k\psi_l\rangle
\end{align}
In order to perform calculations on a quantum computer,  we need to   map  fermionic operators  to qubit operators. We choose Jordan-Wigner transformation to achieve this task  due to its straightforward expression. 

The Jordan-Wigner transformation maps  Eq.~\eqref{eq:ferm_hamiltonian} into a qubit Hamiltonian form
\begin{align}
\label{qubit_hamiltonian}
H=&\sum_{i,\alpha}h_{\alpha}^i\sigma_{\alpha}^i+\sum_{i,j,\alpha,\beta}h_{\alpha\beta}^{ij}\sigma_{\alpha}^{i}\sigma_{\beta}^j+\dots,
\end{align}
where Roman indices $i, j$ denote the qubit on which the operator acts, and Greek indices $\alpha, \beta$ refer to  the type of Pauli operators, i.e.,  $\sigma^i_{x}$ means Pauli matrix $\sigma_{x}$ acting on a  qubit at site $i$. Apparently, $H$ in Eq.~\eqref{eq:ferm_hamiltonian} is a linear combination of unitary Pauli matrices. The  methods used in this paper finding  the molecular ground-state and its energy are all based on it.

In this work, we present the FQE to find the molecular ground-state energy by gradient descent iterations. Gradient descent is one of the most fundamental ways for  optimization, that looks for the target  energy value  along the direction of the steepest descent.  Here it is performed in a quantum computer with the help of linear combination of unitary operators. We analyse the relationships  between the gradient descent  iteration depth and the precision of the ground-state  energy. The explicit quantum circuit to implement the algorithm is constructed. As illustrative examples,  the ground-state energies and electronic structures of four molecules, H$_2$, LiH, H$_2$O and NH$_3$ are presented. Taking H$_2$O and NH$_3$ as examples, a comparison  between the FQE and VQE, a representative hybrid method, is given. FQE can be accelerated further by harnessing perturbation theory in chemical precision. Finally, we analyse the computation complexity  of FQE and summarize the results.

\subsection{Quantum Gradient Descent Iteration}
The classical gradient descent algorithm is usually employed to obtain the minimum of an target function $f(\vecX)$. One starts from an initial point $\vecX^{(0)}={x_1^{0},x_2^{0},\dots,x_N^{0}} \in \mathbb{R}^N$, then moves to the next point along the direction of  the gradient of the  target function, namely
\begin{equation}
\vecX^{(t+1)} = \vecX^{(t)} -\gamma_0\nabla f(\vecX^{(t)}),
\label{Eq:graddesc}
\end{equation}
where $ \gamma_0 $ is a positive learning rate that determines the step size of the iteration.  In  searching the minimum energy of a  Hamiltonian, the target function can be expressed as  a quadratic optimization problem in the form, $f(\vecX)=\vecX^T \matH \vecX$.  At  point $\vecX$, the gradient operator of the objective function can be expressed as
\begin{equation}
\nabla f(\vecX) =  2\matH \vecX.
\label{Eq:deriv}
\end{equation}
Then,  the gradient descent iteration can be  regarded as an evolution of $\vecX$ under operator $\opH$,
\begin{equation}\label{D1}
|\vecX^{(t+1)}\rangle= \left ( |\vecX^{(t)}\rangle -\gamma \matH |\vecX^{(t)}\rangle \right),
\end{equation}
where $\gamma_0$ is redefined as $\gamma=2\gamma_0$.
 In quantum gradient descent, vector $\vecX $  is replaced  by quantum state $\ket{\vecX}=\sum_j x_{j}\ket{j}/\|\vecX\|$, where $x_j$ is the $j$-th elements of the vector, $\ket{j}$ is the $N$-dimensional computational basis, and $ \|\vecX\|$ is the modulus of vector $\vecX $.
Denoting  $\matH^{g}=\matI-\gamma \matH$ and it can be expressed as 
\begin{equation}\label{LCU}
    \matH^{g}=\sum_{i=1}^{M}\beta_{i}\matH^{g}_{i},   
\end{equation}
where  $M$ is the number of Pauli product terms  in  $\matH^{g} $. 
 Then the gradient descent process can be rewritten as 
\begin{equation}\label{D2}
|\vecX^{(t+1)}\rangle=\matH^{g} |\vecX^{(t)}\rangle=\sum_{i=1}^{M}\beta_i\matH^{g}_{i}|\vecX^{(t)}\rangle,
\end{equation}
where $\matH^{g}$ is  a linear combination of unitary operators(LCU) which  was proposed in \cite{gui2006general} in designing quantum algorithms  and studied extensively  \cite{gudder2007mathematical,gui2008duality,gui2009allowable,long2011duality,childs2012hamiltonian,berry2015simulating,wei2016duality}. This non-unitary evolution can be implemented in a unitary quantum circuit by adding ancillary qubits that transform it into unitary evolution  in a larger space \cite{cao2010restricted}. The realization of LCU can be viewed as a quantum computer wavefunction passing through M-slits, and operated by a unitary operation in each slit, and then the wavefunctions are combined and the result of the calculation is readout by a measurement \cite{long2011duality}.  We perform the evolution described by Eq.~\eqref{D2} with the following four steps. 

Wave division: The register is  a composite system which contains a work system and an ancillary register. Firstly, the initial point $\vecX = (x_1, \dots, x_N)^T$ is efficiently mapped as an initial state $|\vecx^{(t)}\rangle$ of the work system. In quantum chemistry, Hartree-Fock (HF) product state is usually used as an initial state.  And the ancillary register is initialized from $ |0\rangle^{m} $, where $m=\text{log}_{2}M $, to a specific superposition state $|\psi_s\rangle$,
\begin{eqnarray}\label{Super}
|\psi_s\rangle=\frac{1}{\mathbb{C}} \sum_{i=0}^{M-1}\beta_{i}|i\rangle
\end{eqnarray}
where $\mathbb{C}=\sqrt{\sum_{i=0}^{M-1} \beta_{i}^{2}} $ is a normalization constant and $|i\rangle$ is the computational basis. This is equivalent to let the state $|\vecx^{(t)}\rangle$ pass through M-slits. $\beta_{i}$ is a factor describing the properties of the slit, which is determined by the forms of the Hamiltonian in Eq.~\eqref{LCU}. This can be done by the initialization algorithm in \cite{long2001efficient}. 
Moreover, the quantum random access memory (qRAM) approach can be used to prepare $|\vecx^{(t)}\rangle$ and $|\psi_s\rangle$, which consume $O(\log N)$ and $O(\log M)$ basic steps or gates respectively after qRAM cell is established. We denote the whole state of the composite system  as $|\Phi\rangle =|\psi_s\rangle |\vecx^{(t)}\rangle$.

Entanglement: Then, a series of  ancillary system controlled operations $ \sum_{i=0}^{M-1}|i\rangle \langle i|\otimes H^{g}_{i} $ are implemented on the work qubits. The work qubits and the ancilla register are now entangled , and the  state is transformed into
 \begin{align}\label{entan}
|\Phi\rangle \rightarrow \frac{1}{\mathbb{C}}\big(\sum_{i=0}^{M-1} \beta_{i}|i\rangle H^{g}_{i} |\vecx^{(t)}\rangle\big).
\end{align}
The corresponding physical picture is that different unitary operations are implemented simultaneously  in different subspaces, corresponding to different slits.

Wave combination : We perform $m$ Hadamad gates on ancillary register to combine all the wavefunctions from the $M$ different subspaces. We merely focus on the component in a subspace where the ancillary system is in state $|0\rangle$. The state of the whole system in this subspace is
 \begin{equation}
|\Phi_0\rangle = \frac{1}{\mathbb{C}\sqrt{2^{m}}
}\big(|0\rangle\sum_{i=0}^{M-1} \beta_{i} H^{g}_{i} |\vecx^{(t)}\rangle\big).\label{comb}
\end{equation}
Measurement: Then, we measure the ancillary register. If we obtain $|0\rangle$, our algorithm succeeds and we obtain the state $\frac{1}{\mathbb{C}\sqrt{2^{m}}}\big(|0\rangle\sum_{i=0}^{M-1} \beta_{i} H^{g}_{i} |\vecx^{(t)}\rangle\big)$, where the work system is in  $|\vecx^{(t+1)}\rangle=\matH^{g} |\vecx^{(t)}\rangle$. And then this will be used as input  for the next iteration in the quantum gradient descent process. The probability of obtaining $|0\rangle$ for the  state  is 
 $$P_{s}=\parallel \matH^{g}|\vecx^{(t)}\rangle\parallel^{2}/{\mathbb{C}^{2}M}.$$ The successful probability after $n$  measurements is  $ 1-(1-\frac{\parallel\matH^{g}|\vecx^{(t)}\rangle\parallel^{2}}{\mathbb{C}^{2}M})^{n}$, which is an exponential function of $n$. The number of measurements is   ${\mathbb{C}^{2}M}/\parallel\matH^{g}|\vecx^{(t)}\rangle\parallel^{2} $. The measurement complexity will grow exponentially with respect to the number of iteration steps \cite{rebentrost2019quantum}.
Alternatively, one can use the  oblivious amplitude amplification \cite{berry2015simulating} to amplify the amplitude of the desired term (ancillary qubits in state $|0\rangle$) up to a deterministic order with $O(\sqrt{M})$ repetitions before the measurement.  Then, the measurement complexity will be  the product of iteration depth $k$ and $O(\sqrt{M})$, linearly dependent on  the number of iteration steps.
After obtaining $|0\rangle$, we can continue  the gradient descent process  by repeating the above four steps, with $|\vecx^{(t)}\rangle$ repalced by $|\vecx^{(t+1)}\rangle$ in Wave-division step. 
We  can  pre-set  a threshold defined as $\varepsilon=|\langle\vecx_{t} |H|\vecx_{t}\rangle-\langle\vecx_{t+1} |H|\vecx_{t+1}\rangle|/\langle\vecx_{t} |H|\vecx_{t}\rangle$ as criterion for stopping the iteration. Thus, we  judge if the iterated state satisfies criterion by measuring the expectation value of  Hamiltonian around the expected number of iteration, which is easier  than constructing the tomography.  If the next iterative state $|\vecx^{(t+1)}\rangle$ does not hit our pre-set threshold, this output $|\vecx^{(t+1)}\rangle$ will be regarded as the new input state $|\vecx^{(t)}\rangle$ and run the next iteration. Otherwise, the iteration can be terminated and the state $|\vecx^{(t+1)}\rangle$ is the final result  $|\vecx_{f}\rangle$, as one good  approximation of the ground state. The ground state energy can be calculated by $\langle\vecx_{f} |H|\vecx_{f}\rangle$.
\begin{figure}
	\centering
	\includegraphics[width=\linewidth]{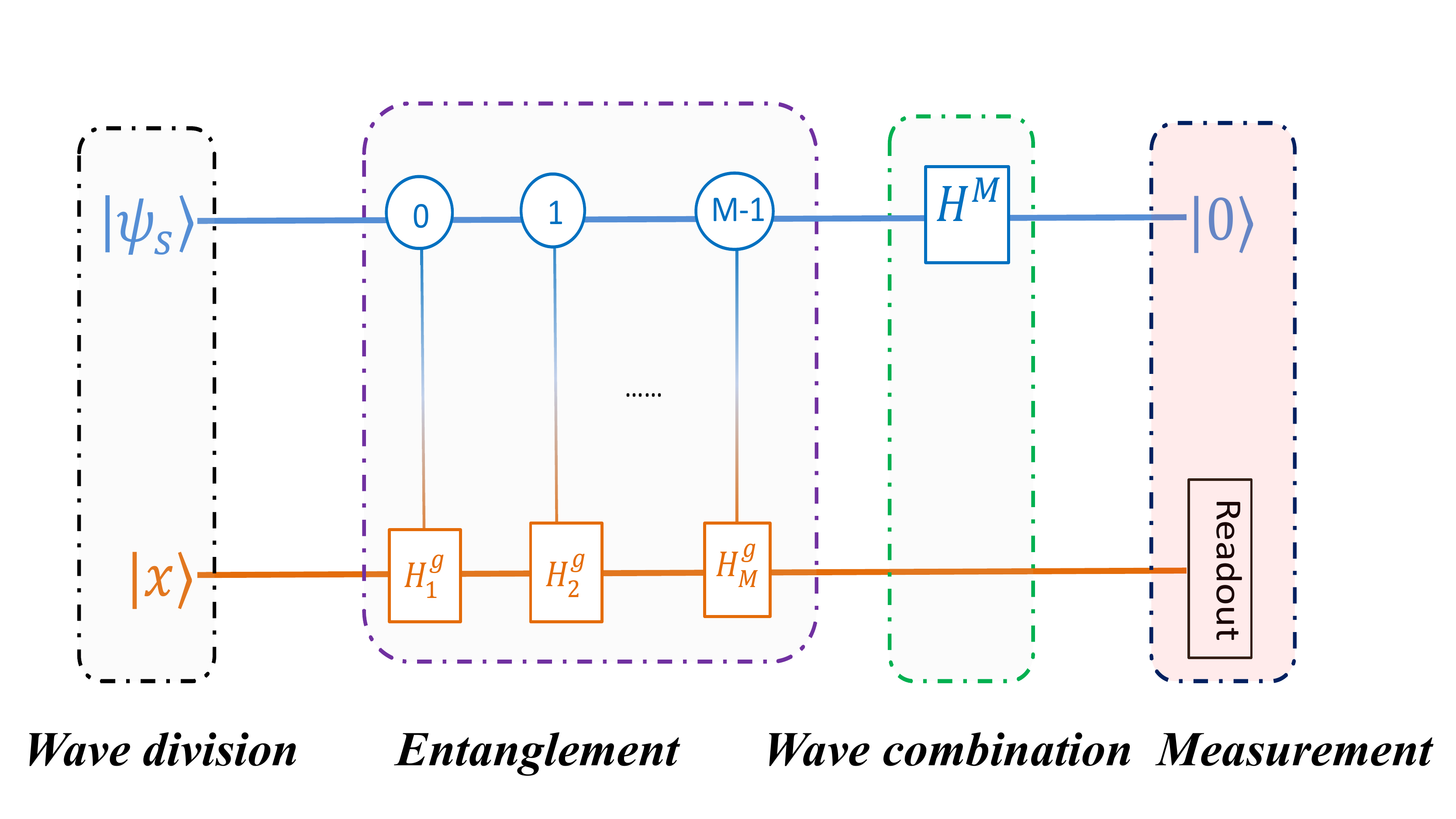}
	\caption{ \footnotesize{Quantum circuit for gradient descent.   $|\vecx\rangle$ and  $|\psi_s\rangle$ denote the initial state of the work system and ancilla syetem respectively. The controlled operations acted on work system are $ \sum_{i=0}^{M-1}|i\rangle \langle i|\otimes H^{g}_{i} $. $H^M$ denotes $m=\text{log}_{2}M $ number Hadamard gates.At the end of the circuit, we measure the final state of the ancilla registers. If all ancilla qubits are $\ket{0}$, the work systerm  collapses into state $|\vecx^{(t+1)}\rangle$. } } \label{cir}
	\end{figure}

Measuring the expectation values during the iteration procedure will destroy the  state of the work system, stopping the quantum gradient descent process.
So,  determing the iteration depth $k$ in advance is essential. After $k$ times iterations, the approximation error is limited to (ignoring constants)
\begin{eqnarray*} 
\epsilon \leq O\left((\frac{1-\gamma\lambda_2}{1-\gamma\lambda_1})^kN\right), 
\end{eqnarray*}
where $ \lambda_1$ and $\lambda_2$ are the two largest absolute values of the eigenvalues of Hamiltonian $\matH$ (see  Supplemental Material  for proof). The iteration depth 
\begin{equation}
    k=O(\log\frac{N}{\epsilon})
    \label{depth}
    \end{equation}
is logarithmically dependent on the system size and the inverse of precision. The algorithm may be terminated at a point with a pre-set  precision $\epsilon$. It can be seen that the choise of $\gamma$ has little  impact on converge rate when $\gamma$ is  large. This makes this algorithm very robust to this parameter. The rate of convergence primarily depends upon the ratio of $ \lambda_1$ and $\lambda_2$.
The gap between the iterative result and the ground state depends on the choice of initial point. If we choose an ansatz state with a large overlap with the exact  ground state, the iterative process will converge to the the ground state in fewer iterations. Usually,  the mean-field state which represents a good classical approximation to the ground state of Hamiltonian H, such as a Hartree-Fock (HF) product state, is chosen as an initial state. Compared to VQE, FQE does not need to make measurements of the expectation values of Hamiltonian during each iteration procedure and this substantially reduces the  computation resources.

\subsection{Perturbation Theory}
The FQE involves multi-time iterations to obtain an accurate result, which is difficult to implement in the present-day quantum computer hardware. Here, we present an  approximate method to find the ground state and its energy by using the gradient descent algorithm and  perturbation theory.
Perturbation theory is widely used and plays an important role in describing real quantum systems, because it is impossible  to find exact solutions to the Schr\"{o}dinger equation for Hamiltonians even with moderate complexity.  The  Hamiltonian described by Eq.~\eqref{qubit_hamiltonian} can be divided into two classes,  $\matH_0$ and $\matH'$. $\matH_0$  consists of a set of Pauli terms containing only  $\sigma^i_{\alpha=z}$ and the identity matrices,  and Pauli terms $\sigma^i_{\alpha=x, y}$ belong to $\matH'$. 
$\matH_0$ is a diagonal matrix with exact solutions, that can be regarded as a simple system. $\matH'$ usually  is smaller compared to $\matH_0$, and is treated as a ``perturbing" Hamiltonian.  The energy levels and eigenstates associated with the perturbed system can be expressed as ``corrections" to those of the unperturbed system.  We begin with the time-independent Schr\"{o}dinger equation: 
\begin{align}
\matH|\psi_{n} \rangle = (\matH_0+\matH') |{\psi}_{n} \rangle = E_{n}|{\psi}_{n} \rangle,
\label{comb}
\end{align}
where $E_{n}$ and $|\psi_{n} \rangle$ are the n-th  energy and eigenstate respectively. Unperturbed Hamiltonian  $\matH_0$, satisfies the time-independent Schr\"{o}dinger equation: $\matH_{0}|n \rangle=E_{n}^{(0)}|n\rangle$. Our goal is to express $ E_{n}$ and $|\psi_{n} \rangle$ in terms of $E_{n}^{0}$ and $|n\rangle$. Denote the expectation value of $\matH'$ as $\matH'_{nn}=\langle{n}| \matH'|{n} \rangle$, and it is easily to see that $\langle{n}|\matH'|{n} \rangle$ is zero  because $\matH'$ only contains Pauli terms $\sigma^i_{\alpha=x, y}$. In the first order approximation, the energies and eigenstates are expressed as
\begin{align}
E_{n} &= E_{n}^{(0)},\\
|\psi_{n} \rangle&=|{n} \rangle -\sum_{m\ne n}\frac{\matH'_{mn}}{E_{m}^{(0)}-E_{n}^{(0)}}|m \rangle.
\label{first}
\end{align}
To second-order approximation, they are
\begin{align}\label{second}
E_{n} &= E_{n}^{(0)}+\sum_{m\ne n}\frac{|\matH'_{mn}|^2}{E_{m}^{(0)}-E_{n}^{(0)}},\\
|\psi_{n} \rangle&=|{n} \rangle -\sum_{m\ne n}\frac{\matH'_{mn}}{E_{m}^{(0)}-E_{n}^{(0)}}|m \rangle -\frac{1}{2}\sum_{m\ne n}\frac{|\matH'_{mn}|^2}{(E_{m}^{(0)}-E_{n}^{(0)})^2}|n \rangle,\\\nonumber
&+\sum_{m\ne n} \left[\sum_{k\ne n}\frac{\matH'_{mn}\matH'_{kn}}{(E_{m}^{(0)}-E_{n}^{(0)})(E_{k}^{(0)}-E_{n}^{(0)})}\right]|m \rangle.
 \end{align}
The matrix elements in the first and second-order approximations can be obtained by one  iteration of the quantum circuit in Fig.\Ref{cir}.  Here, we let $\matH'$ be equal to $\matH^g$.
Explicitly, the  first order approximation only involves $ \matH'_{mn}$,  a  series of transition probabilities of the state after $\matH'$ implemented on state $|{n} \rangle$,  and they can be obtained by performing the quantum circuit of Fig.\Ref{cir} directly. For the  second order approximation, matrix elements such as value $|\matH'_{mn}|^2$ and $\matH'_{mn}\matH'_{kn}$, can be calculated by $ \matH'_{mn}$. Then, the  approximate ground energy and ground state up to second-order are obtained. We will show the performance of FQE and perturbation theory in  next section.

\section{results}
\subsection{Calculations of Four Molecules}
\begin{figure}
\centering
\includegraphics[width=\linewidth]{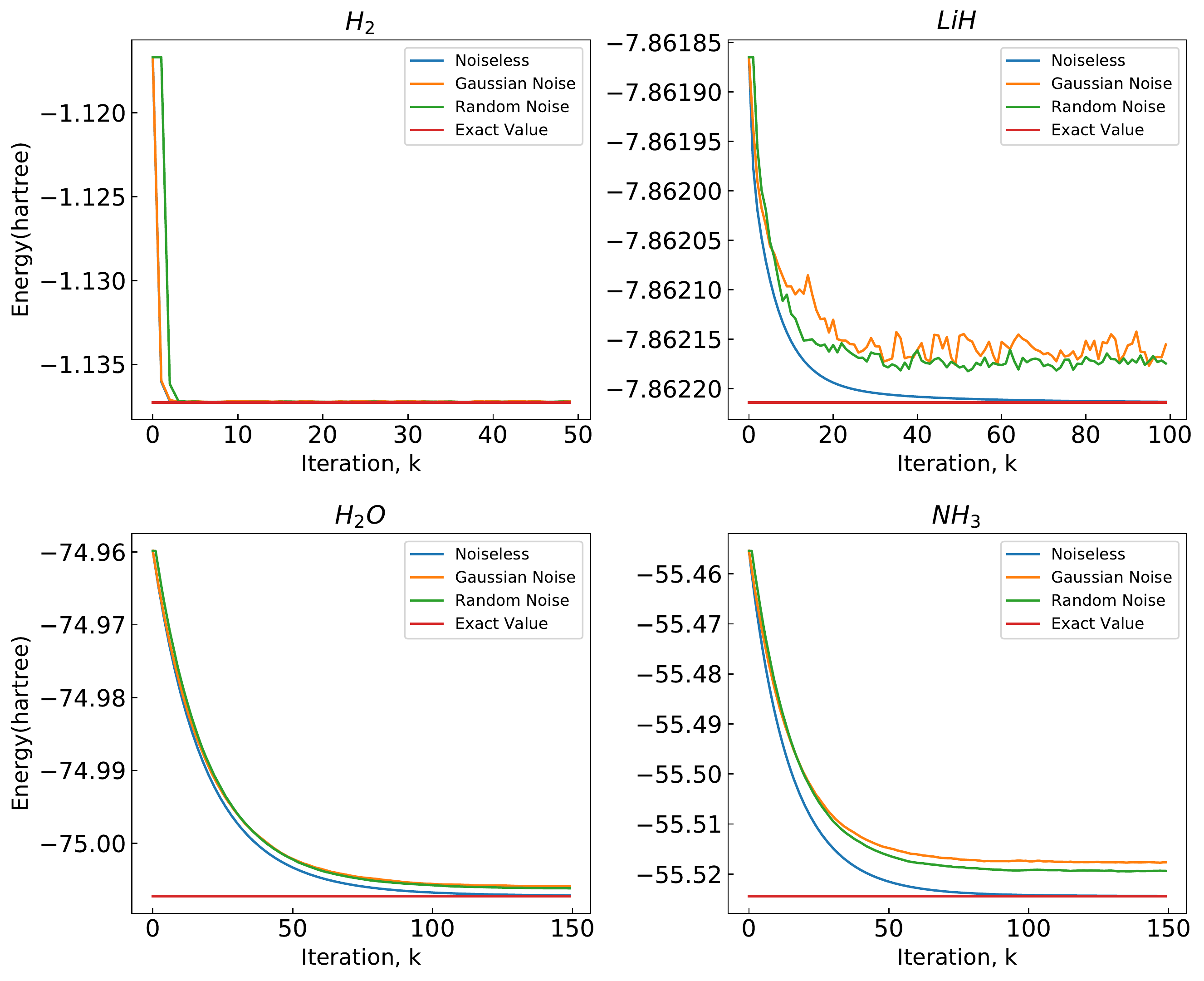}
\caption{ \footnotesize{(a), (b), (c) and (d) show the convergence  to ground state energies by FQE for H$_{2}$, LiH, H$_{2}$O and NH$_{3}$ molecules respectively.  The numerical simulations are carried out with  fixed interatomic distance. The exact value corresponding to Hamiltonian diagonalization energy (red line). The initial state is chosen as Hartree-Fock product state in all four cases. The final values of  the lines for exact ground state energy (red line) and for the three  iteration results, noiseless case (blue line),  random noisy case (green line) and Gaussian noisy case (orange line). }} \label{iter}
\end{figure}

To demonstrate the feasibility of this FQE with gradient descent iteration,  we carried out calculations on the ground state energy of  H$_{2}$, LiH diatomic molecules, and two relatively complex molecules H$_2$O and NH$_3$.  We used a common molecular basis set, the minimal STO-3G basis. Via Jordan-Wigner transformation, the qubit-Hamiltonians of these molecules are obtained.   The Hamiltonians of H$_{2}$, LiH, H$_{2}$O and NH$_{3}$ contain 15 , 118, 252, and 3382 Pauli matrix product terms respectively. The dimensions of the Hamiltonians of H$_{2}$, LiH, H$_{2}$O and NH$_{3}$ are 16 , 64, 4096, and 16384 respectively, which corresponds 4, 6, 12, 14 number of qubits respectively. In all  four simulations, the work system was initialized to the HF state $|\vecx_{h}\rangle$ and  the learning rate is chosen as  $\gamma=1$. As shown in Fig.\Ref{iter}, after about 120 iterations, the molecular energy of  H$_{2}$O  converges to -74.94 a.u, only $0.0013346\%$ discrepancy  with respect to the exact value of -74.93 a.u. obtained via Hamiltonian diagonalization. The NH$_{3}$ calculation yields (-55.525 a.u.) after 80 iterations, matched very well with the diagonalization (-55.526 a.u.).  For the study of atomic  molecular structures and chemical reactions, these results are sufficiently  accurate. For more complex basis set STO-6G, the results are about the same, and  the details are given in  Supplemental Material. 
The converge rates of the four molecules depend on the system size and the ratio of the two largest absolute eigenvalues of the Hamiltonian $\matH$, which are consistent with the theoretical analysis above.

\begin{figure}
\centering
\includegraphics[width=\linewidth]{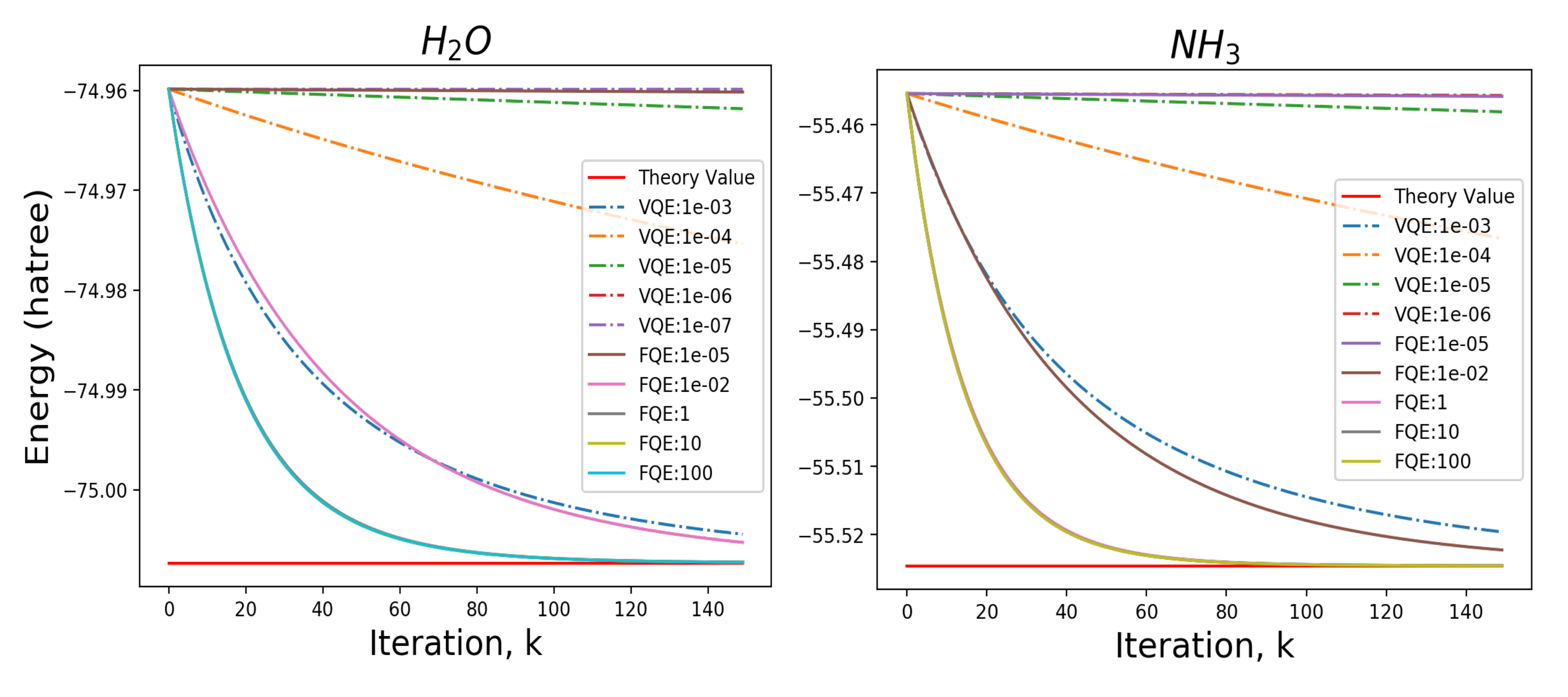}
\caption{ \footnotesize{ The exact comparison of  FQE  and VQE  for searching ground state energy of H$_{2}$O and NH$_{3}$ molecules respectively. The red color lines labled as 'Theory Value' are the exact values of ground state energy.  The labels of the right symbols denote different learning rates. }} \label{compare}
\end{figure}

We also studied the infulence of noises which is also shown in the  Fig.\Ref{iter}. The noise term is chosen  the form of $ \sum_{i=1}^{N}\delta\alpha_i\sigma_{z}$ , added to the Hamiltonian  to simulate decoherence. Then we add a term $|\delta\vec x\rangle$ on the iterative state $|\vecx^{k}\rangle$ to simulate measurement error and renormalize the iterative state as $|\vecx^{(k)}\rangle\rightarrow(|\vecx^{(k)}\rangle+|\delta\vec x\rangle)/\parallel|\vecx^{(k)}\rangle+|\delta\vec x\rangle\parallel$.
We set a random noise ( amplitude $0.01$) and a Gaussian noise ($\mu=0,\sigma=0.01/3$) for H$_{2}$ and LiH. For H$_{2}$O and NH$_{3}$, we choose a random noise (amplitude $0.02$) and a Gaussian noise ($\mu=0,\sigma=0.02/3$).  The results still converge to the exact values in chemical precision ($1.6\times 10^{-3}$ a.u). This indicates that our method is robust to certain type of noise, which is important in the implementation of quantum simulation on near term quantum devices.  For more noisy situations, see Supplemental Material for details, where the  parameters of noise are 10 times of the above values. The convergence deteriorates and some oscillations accur as the number of iterations increases. 
 
In Fig.\Ref{compare}, a comparison with VQE is shown for H$_2$O and NH$_3$.   In VQE calculation, the initial state $\ket{\vecx_0}$ is mapped  to an ansatz state by a parameterized unitary operation $\ket{\vecx({\vec \theta})}=U(\vec \theta)\ket{\vecx_0}$.  VQE solves for the parameter vector $\vec \theta$ with a classical optimization routine. Here we adopt the standard gradient descent method as  the classical optimizer in VQE. The  parameter  is updated by $\vec \theta\rightarrow \vec \theta-\gamma\frac{f(\vec \theta+\Delta\vec \theta)-f(\vec \theta)}{\Delta\vec \theta}$. We performed numerical simulations of VQE for the two molecules.  When  the  learning rate $\gamma\geq 10^{-3}$, VQE does not converge to the ground state. So, in order to compare with each other, we choose the proper learning rates in two  methods seperately. In both cases, the initial ansatz state is prepared as the HF product state. In H$_2$O and NH$_3$, VQE converges most fast with the learning rate $\gamma= 10^{-3}$. FQE converges more and more fast with larger and larger  learning rate   until a fixed speed is reached . As shown in Fig.\Ref{compare}, FQE generally converges faster than VQE and the advantage will be more obvious in  complex molecules.

\begin{figure}
\centering
\includegraphics[width=\linewidth]{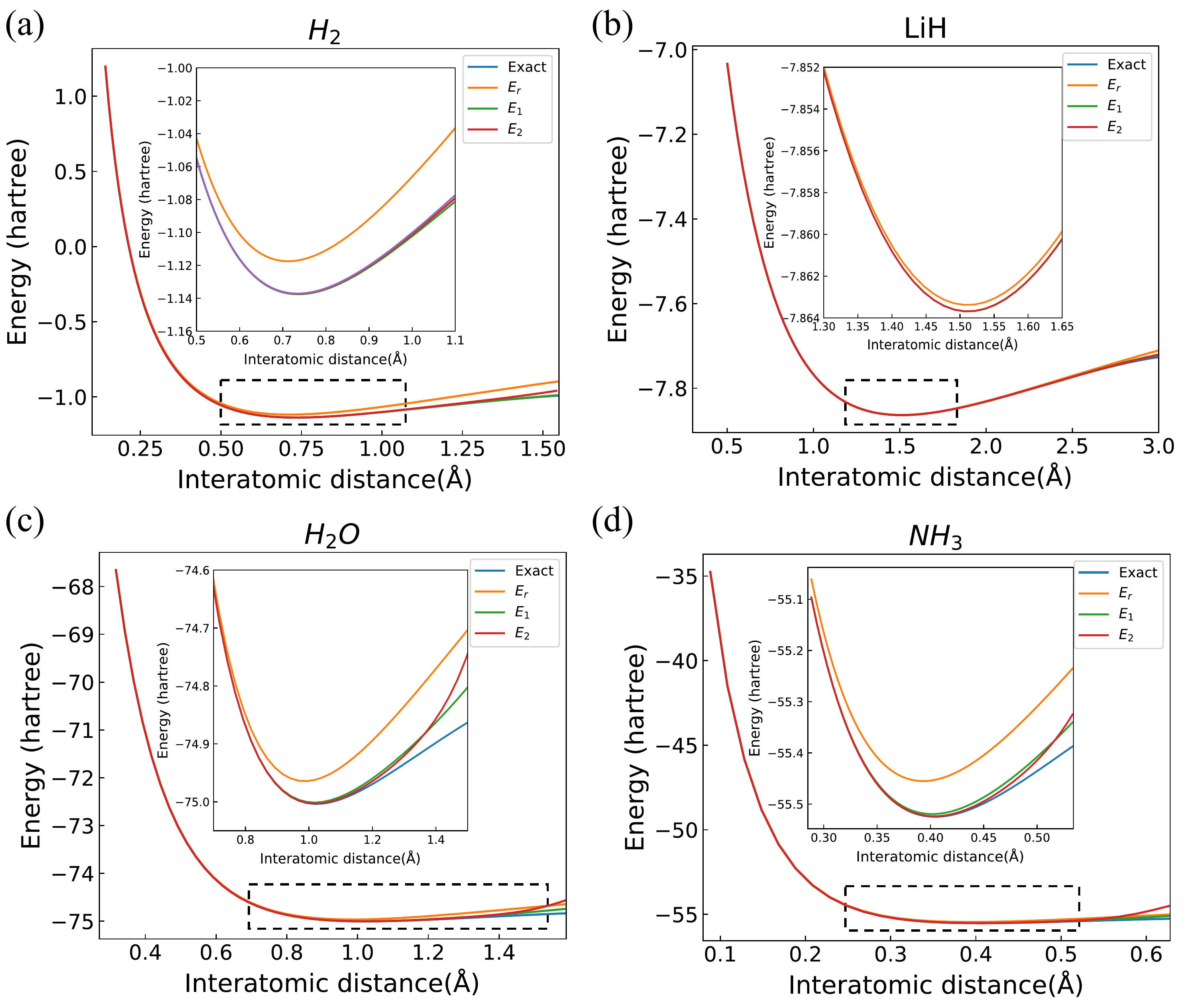}
\caption{ \footnotesize{Theory results (blue lines), zero-order (orange lines), first-order (green lines) and second-order (red lines) energy plots of outcomes from numerical simulations, for several interatomic distances for H$_{2}$, LiH, H$_{2}$O (between the oxygen atom and one hydrogen atom) and NH$_{3}$ (between the nitrogen atom and the plane formed by the three hydrogen atoms). }} \label{pertur}
\end{figure}

The above examples are calculated in fixed interatomic distance of the molecules. If we want to calculate  the  interatomic distance corresponding to the most stable structure, the variation of interatomic distances is necessary.
 In Fig.\Ref{pertur}, four examples are given to illustrate the performance of perturbation theory. To obtain the potential-energy surfaces for  H$_{2}$, LiH, H$_2$O and  NH$_3$ molecules, we  studied  the dependence of  ground-state energy of their molecules on the  variating interatomic distances, between the two atoms in H$_{2}$, LiH, and the distance between the oxygen atom and one hydrogen atom (the two hydrogen atoms are symmetric with respect to the oxygen atom) in H$_2$O, and the  distance between the nitrogen atom and the plane formed by the three hydrogen atoms  in NH$_3$. The lowest energy in potential-energy surfaces corresponds to  the most stable structure of the molecules. As shown in the picture, the ground-state energy of each molecule calculated under the second order approximation are already quite close to their exact values, which is obtained from  Hamiltonian diagonalizations. The energy values up to second-order correction are compared with their exact values at the most stable interatomic distance corresponding to the lowest energy in Table.1. It can be seen that the second order approximation has already given results in chemical precision.

\begin{table*}
\centering
\begin{tabular}{|l|c|c|c|c|c|}
\hline
\diagbox{Distance($\mathrm{\AA}$)}{Energy value(au)} & exact value & zero-order value & first-order value & second-order value  \\
\hline
~~H$_2$(0.7314) & -1.1373 & -1.1171 & -1.1372 & -1.1372 \\
\hline
~~LiH(1.5065)  & -7.8637& -7.8634 &-7.8637 &-7.8637 \\
\hline
~~H$_2$O(1.0812) & -75.0038 & -74.9622 & 75.0013 &75.0032  \\
\hline
~~NH$_3$(0.4033)  & -55.5247 & -55.4530 & -55.5193 &-55.5237  \\
\hline
\end{tabular}
\caption{\footnotesize{ Energy values  calculated by perturbation method and  the exact values in the most stable distance corresponding to the lowest ground energy. } } \label{results}
\end{table*}

\subsection{Analysis of Computational Complexity }
Here we analyze the complexity of our algorithm. Usually, a quantum algorithm complexity involves two aspects: qubit resources and gate complexity. For qubit resources,  the number of ancilla qubits is $ \text{log}M$, where $M$ is the number of Pauli terms in qubit form Hamiltonian. For gate complexity, the ``Wave division''  part needs $\mathcal{O}(\text{log}N+\text{log}M)$ basic steps for state preparation. The dominate factor is the number of controlled operations in ``Entanglement''	 part in Fig. \Ref{cir}.  Controlled $\matH^{g}_{i}$ can be decomposed into $\mathcal{O}(M\text{log}M\text{log}N)$ basic gates \cite{xin2017quantum,wei2018efficient}. The ``Wave combination '' part just comprises  $ \text{log}M$ Hadamard gates. Totally, FQE requires in each iteration  about $\mathcal{O}(M\text{log}M\text{log}N)$ basic gates for implementation. If the wavefunction is expressed by $O(N)$ Gaussian orbitals, fermion Hamiltonians contain $ O(N^4)$ second-quantized terms, consequently the  qubit Hamiltonians have $ M=O(N^4)$ Pauli terms. The qubit resource and gate complexity can be reduced to  $O(N)$ and $O(N^4)$ respectively. In some applications, the perturbation theory only requires one  iteration, and an approximate result in chemical precision can be obtained.  

 \section{Summary}
 An efficient quantum  algorithm, Full Quantum Eigensolver (FQE), for calculating the ground state wavefunction and the ground energy using gradient descent (FQE) was proposed, and numerical simulations are performed for four molecules. In FQE, the complexity of  basic gates operations is polylogarithmical to the number of single-electron atomic orbitals. It achieves an  exponential speedup compared with its classical counterparts. It has been shown that FQE is robust against noises of reasonable strengths. For very noisy situations that do not allow many iterations, FQE can be combined with perturbation theory that give the ground state and its energy in chemical precision with one time iteration. FQE is exceptionally useful in  quantum chemistry  simulation, especially for the near-term NISQ applications. FQE is a full quantum algorithm, not only applicable for NISQ computers, but directly applicable for future large-scale fault-tolerant quantum computers.

\bigskip\noindent
\textbf{Acknowledgements}\\
This research was supported by National Basic Research Program of China. We gratefully acknowledges support from 
the National Natural Science Foundation of China under Grants No. 11974205, and No. 11774197. The National Key Research and  Development Program of China (2017YFA0303700); The Key Research and  Development Program of Guangdong province (2018B030325002); Beijing Advanced Innovation Center for Future Chip (ICFC).
\\
\textbf{Author contributions}\\
S.J.W conceived the algorithm. H.L performed classical simulations. G.L.L initialized LCU scheme.  All authors contributed to the discussion of results and writing of the manuscript.\\
\textbf{Competing interests}\\
The authors declare no competing interests.\\
\textbf{Data availability}\\
The data that support the findings of this study are available from the corresponding authors on reasonable request.\\
\bibliographystyle{unsrt}

\bibliography{QM}
\section{ Supplemental Material}
\subsection{Error estimation and  iteration complexity }

We analyse  FQE's convergence and estimate the approximation error and iteration complexity \cite{panju2011iterative}.
Define $|\psi_i\rangle$ as an normalized eigenvector for $ \matH\in \mathbb{R}^{n \times n}$ with eigenvalue $\lambda_i$,  $\matH|\psi_i\rangle = \lambda_i |\psi_i\rangle$. Suppose that $\matH$ has  real and distinct eigenvalues set $\{\lambda_i\}$ such that $\left|\lambda_1\right| > \left|\lambda_2\right| > \ldots > \left|\lambda_n\right|$. We can express an arbitrary state $|\psi\rangle$ as a linear combination of the eigenvectors of $H$: $$|\psi\rangle = a_1|\psi_1\rangle + \ldots + a_n|\psi_n\rangle.$$ Define  matrix $\matH^{g}=\matI-\gamma \matH$ and perform it on $|\psi\rangle$, we have 

$$\matH^{g}|\psi\rangle = a_1(1-\lambda_1)|\psi_1\rangle + a_2(1-\lambda_2) |\psi_2\rangle +\ldots + a_n(1-\lambda_n) |\psi_n\rangle$$
and so
\begin{eqnarray*}
	(\matH^{g})^k |\psi\rangle & = & a_1(1-\gamma\lambda_1)^k|\psi_1\rangle + c_2(1-\gamma\lambda_2)^k |\psi_2\rangle\\
	 & + & \ldots + c_n(1-\gamma\lambda_n)^k|\psi_n\rangle  \\
	& = &  (1-\gamma\lambda_1)^k \left( a_1|\psi_1\rangle + a_2\left( \frac{(1-\gamma\lambda_2)}{(1-\gamma\lambda_1)}\right)^k|\psi_2\rangle\right. \\
	 & + & \left.\ldots + a_n\left( \frac{(1-\gamma\lambda_n)}{(1-\gamma\lambda_1)}\right)^k|\psi_n\rangle \right) \\
\end{eqnarray*}

Since the eigenvalues are assumed to be real, distinct, and ordered by decreasing magnitude, it follows that for all $i = 2, \ldots, n$, 
$$\lim_{k\rightarrow \infty}\left( \frac{(1-\gamma\lambda_i)}{(1-\gamma\lambda_1)}\right)^k =0.$$
 In the case of molecule Hamiltonian $H$, all of the eigenvalues are less than $0$. Note that  $|\psi_1\rangle$ is the ground state with ground energy $\lambda_1$. As $k$ increases, $(H^{g})^k|\psi\rangle$ approaches the  state $a_1(1-\lambda_1)^k|\psi_1\rangle$, and thus for large value of $k$,
$$|\psi_1\rangle \approx \frac{(H^{g})^k|\psi\rangle}{\sqrt{\langle\psi|(H^{g})^{2k}|\psi\rangle}}.$$ 
The  approximation error
\begin{eqnarray*} 
\epsilon &=&\frac{\langle\psi|(H^{g})^kH(H^{g})^k|\psi\rangle}{\langle\psi|(H^{g})^k(H^{g})^k|\psi\rangle}-\lambda_1\\
&=&\frac{\sum_{i=2}^{n} a_i\lambda_i\left( \frac{(1-\gamma\lambda_i)}{(1-\gamma\lambda_1)}\right)^k}{\sum_{i=1}^{n} a_i\left( \frac{(1-\gamma\lambda_i)}{(1-\gamma\lambda_1)}\right)^k}\\
&\leq&\left(\frac{1-\gamma\lambda_2}{1-\gamma\lambda_1}\right)^k\frac{(n-1)a_2\lambda_2}{a_1}, 
\end{eqnarray*}
which  decreases exponentially in the iteration depth $k$.
If  a good initial state  is chosen so that $a_1$ is large, for instance HF state, $\epsilon$ will be small in early iterations.
With iteration increasing, the  state gets closer and closer to the ground $|\psi_1\rangle$. The algorithm may be terminated at any point with a reasonable accuracy $\epsilon$ to the ground state. 

 The rate of convergence primarily depends upon the ratio of the two eigenvalues of largest absolute value. In the circumstance that the two largest eigenvalues have similar sizes, the convergence will be slow in early stage. That case needs special  attention, and  will not be discussed  here.
 \subsection{ FQE with STO-6G basis as input}
\begin{figure}
 \centering
 \includegraphics[width=\linewidth]{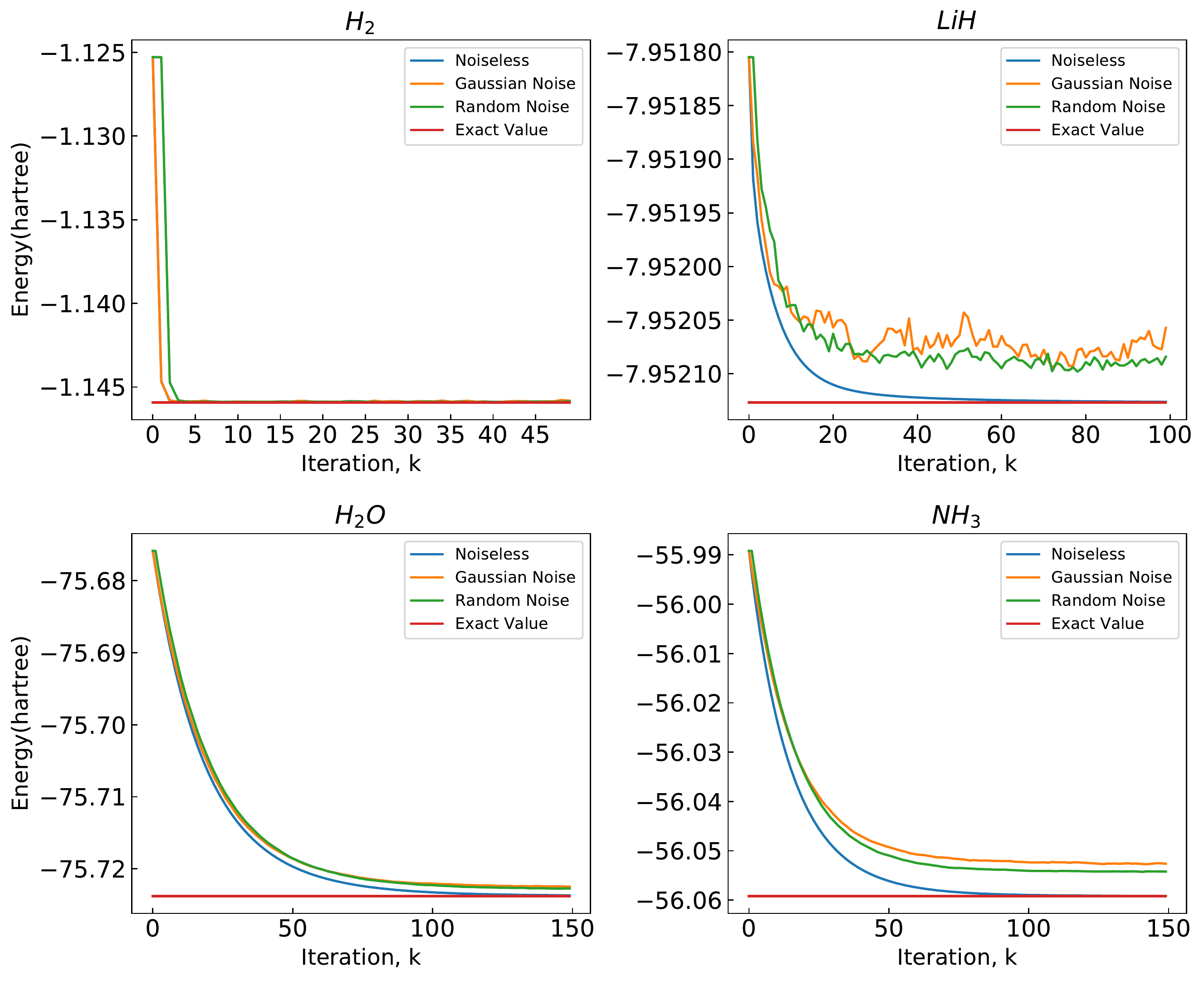}
 \caption{\footnotesize{(a), (b), (c) and (d) show the gradient descent iteration process  for convergence of ground state energy of H$_{2}$, LiH, H$_{2}$O and NH$_{3}$ respectively.  The qubit Hamiltonians of the four molecules are obtained by STO-6G basis, which is more accurate than STO-3G basis. }} \label{6g}
 \end{figure}

 To make our method more plausible, we adopt STO-6G basis sets to generate the qubit Hamiltonians of the four molecules. The noise parameters are as same as the parameters in  the maintext.  The performance of our method is as same as  in STO-3G basis, shown in Fig.\Ref{6g}.
 
 \begin{figure}
	\centering
	\includegraphics[width=\linewidth]{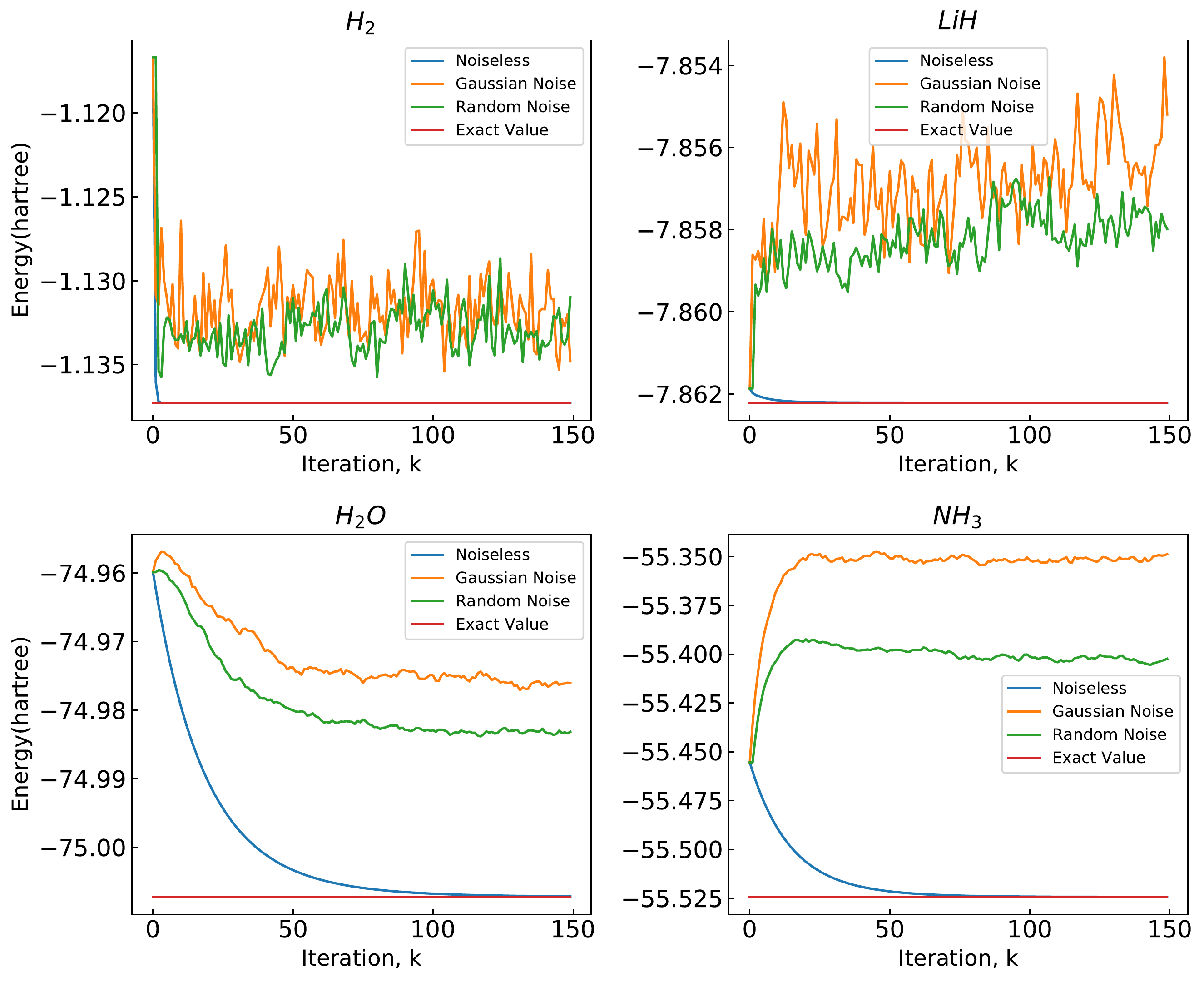}
	\caption{ \footnotesize{Infulence of  large noise on FQE in (a) H$_{2}$, (b) LiH, (c) H$_{2}$O and (d) NH$_{3}$ molecules respectively.  The amplitude of the random noise is 0.1 and the Gaussian noise parameters are $\mu=0,\sigma=0.1/3$. }} \label{big}
	\end{figure}
\subsection{Performance of FQE with large noise}
We show the performance of FQE in large noise situations. As shown in Fig.\Ref{big}, when random noise becomes large, FQE will not converge to the ground state. Sometimes, it  converges to exited energy-levels, such as in H$_{2}$O and NH$_{3}$. In some other situations, FQE  behaves in  a  oscillation manner, such as in H$_{2}$ and LiH.


\end{document}